\documentclass[11pt,twoside]{article}

\usepackage{asp2006}
\usepackage{epsf}

\newcommand{\ggkMgII}{\hbox{{\rm Mg}\kern 0.1em{\sc ii}}}

\begin{document}
\title{Connecting Galaxy Disk and Extended Halo Gas Kinematics}   
\author{G. G. Kacprzak\altaffilmark{1}, C. W. Churchill\altaffilmark{1}, C. C. Steidel\altaffilmark{2}, D. Ceverino\altaffilmark{1},\\ A. A. Klypin\altaffilmark{1}, and M. T. Murphy\altaffilmark{3}}

\altaffiltext{1}{New Mexico State University, Las Cruces, NM 88003}
\altaffiltext{2}{Caltech, Pasadena, CA 91125}
\altaffiltext{3}{Swinburne University of Technology, Hawthorn,
Victoria 3122, Australia}

\begin{abstract}
We have explored the galaxy disk/extended halo gas kinematic
relationship using rotation curves (Keck/ESI) of ten intermediate
redshift galaxies which were selected by {\ggkMgII} halo gas
absorption observed in quasar spectra.  Previous results of six
edge--on galaxies, probed along their major axis, suggest that
observed halo gas velocities are consistent with extended disk--like
halo rotation at galactocentric distances of 25--72~kpc. Using our new
sample, we demonstrate that the gas velocities are by and large not
consistent with being directly coupled to the galaxy kinematics.
Thus, mechanisms other than co--rotation dynamics (i.e., gas inflow,
feedback, galaxy--galaxy interactions, etc.)  must be invoked to
account for the overall observed kinematics of the halo gas.

\noindent In order to better understand the dynamic interaction of the
galaxy/halo/cosmic web environment, we performed similar mock
observations of galaxies and gaseous halos in $\Lambda$--CDM
cosmological simulations. We discuss an example case of a $z=0.92$
galaxy with various orientations probing halo gas at a range of
positions. The gas dynamics inferred using simulated quasar absorption
lines are consistent with observational data.

\end{abstract}

\section{Observations \& Simulations of Extended {\ggkMgII} Halo Gas}

We have extracted rotation curves from ESI spectra of 10 {\ggkMgII}
absorption selected galaxies. The galaxies have a range of inclination
and position angles with respect to the quasar line of sight and a
range of impact parameters between 26--107 kpc. Figure~1 shows an
example case of the quasar field Q0454--220 where $z_{gal}= 0.48382$
and $D= 107$~kpc. The extended {\ggkMgII} absorbing gas aligns with
one side of the galaxy rotation curve. Interestingly, the clouds are
blended together, which is suggestive that the gas exhibits some form
of organized motion (e.g., Figure~1b). In fact, in all but one case,
we find the absorbing gas aligns with one side of the rotation
curves. However, with this diverse sample, {\it disk--like halo
rotation models cannot reproduce the observed halo gas velocity
spread. Thus, disk--like halo rotation does not seem to dominate over
infall and outflow kinematic contributions as was previously
suggested} \citep{s02}.

\begin{figure}[h]
\plotfiddle{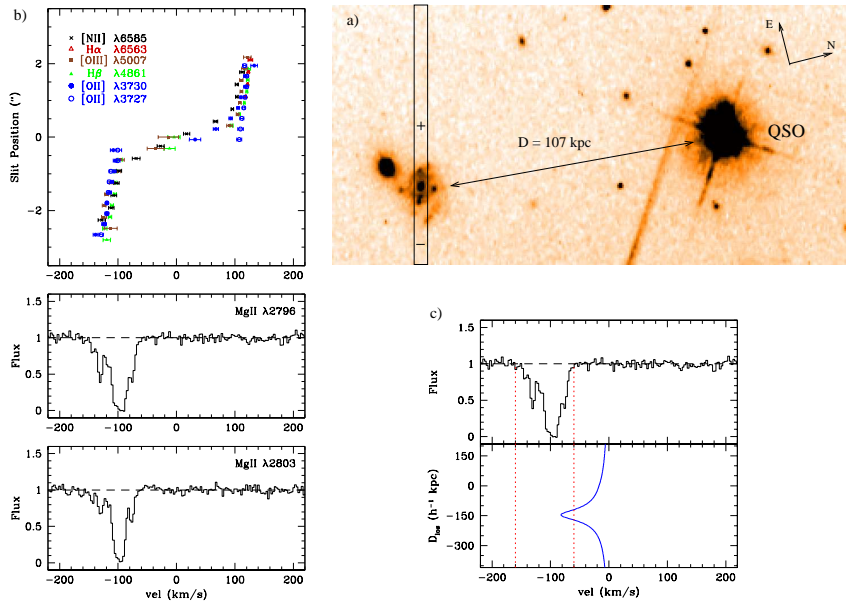}{1.3in}{0.}{41.}{41.}{-165}{-137}
\vglue 1.69in
\caption{--- a) WFPC2 $30\times15''$ image of the quasar field. The
``$+$'' and ``$-$'' indicate positive and negative arcseconds along
the slit relative to the center of the galaxy. --- b) (upper) Emission
line rotation curves of the galaxy obtained with ESI. (lower) HIRES
quasar spectrum of the {\ggkMgII} absorbing gas. --- c) (upper)
{\ggkMgII} 2796 absorption kinematics. (lower) The extended disk model
of halo velocity as a function of position along the line of sight,
where $D_{los}=0$ occurs where the line of sight intersects the
extrapolated disk mid--plane. The lack of velocity overlap between the
model and the observed absorbing gas reveals that disk--like halo
rotation cannot account for the total gas kinematics.}

\end{figure}   

Such discrepancies with the simple models have motivated us to
investigate the connection between disk/halo/IGM gas in cosmological
simulations of galaxy formation.  The simulations are performed using
the Eulerian Gas dynamics plus N--body Adaptive Refinement Tree (ART)
code \citep{k2}.  We ran mock quasar sightlines through the
simulations sampling every 10~kpc in a 80~kpc grid around the
galaxies.  Approximately 30\% of the sightlines, primarily those
probing along the major axes, are consistent with disk--like rotation.
Although it naively appears that the absorption kinematics are
consistent with disk rotation, inspection of the simulations reveals
that the gas does not necessarily co--rotate with the galaxies.  In
70\% of sightlines, simple models of disk rotation fail to reproduce
the {\ggkMgII} absorption velocities and velocity spreads.  Thus, we
find that, statistically, the kinematics of halo gas in cosmological
simulations are not inconsistent with observed halo gas kinematics.
Simulations suggest that stellar winds, SNe, minor
mergers/harassments, and cool filament gas inflows all contribute to
the dynamics of the halo gas. The dynamic contribution of all of these
mechanisms are dependent on the galactocentric distance as well as the
evolutionary state (quiescent or star forming) of the host galaxy.



\end{document}